\documentstyle[12pt,sprocl]{article} 
\newcommand{\nc}{\newcommand}
\nc{\renc}{\renewcommand}

\nc{\half}{{\textstyle{1\over2}}}
\newcommand{\pv}{\mbox{\boldmath $p$}}
\newcommand{\kv}{\mbox{\boldmath $k$}}

\nc{\etal}{\mbox{\it et al. }}
\nc{\ie}{{\it i.e.}}
\nc{\eg}{{\it e.g.}}

\renc{\thefootnote}{\arabic{footnote}}
\nc{\capt}[1]{{\bf Figure.} {\small\sl #1}}

\nc{\eqs}[2]{\mbox{Eqs.~(\ref{#1},\,\ref{#2})}}
\nc{\eq}[1]{\mbox{Eq.~(\ref{#1})}}

\nc{\figs}[2]{\mbox{Figs.~(\ref{#1},\,\ref{#2})}}
\nc{\fig}[1]{\mbox{Fig~.(\ref{#1})}}

\nc{\tag}[1]{\label{#1} \marginpar{{\footnotesize #1}}}
\nc{\mtag}[1]{\label{#1} \mbox{\marginpar{{\footnotesize #1}}}}
\renc{\baselinestretch}{1.2}
\newlength{\overeqskip}
\newlength{\undereqskip}
\nc{\be}[1]{\begin{equation} \mbox{$\label{#1}$}}
\nc{\bea}[1]{\begin{eqnarray} \mbox{$\label{#1}$}}
\nc{\Section}[2]{\section{#2}\label{#1}}
\nc{\Bibitem}[1]{\bibitem{#1}}
\nc{\Label}[1]{\label{#1}}

\nc{\eea}{\vspace{\undereqskip}\end{eqnarray}}
\nc{\ee}{\vspace{\undereqskip}\end{equation}}
\nc{\bdm}{\begin{displaymath}}
\nc{\edm}{\end{displaymath}}
\nc{\dpsty}{\displaystyle}
\nc{\bc}{\begin{center}}
\nc{\ec}{\end{center}}
\nc{\ba}{\begin{array}}
\nc{\ea}{\end{array}}
\nc{\bab}{\begin{abstract}}
\nc{\eab}{\end{abstract}}
\nc{\btab}{\begin{tabular}}
\nc{\etab}{\end{tabular}}
\nc{\bit}{\begin{itemize}}
\nc{\eit}{\end{itemize}}
\nc{\ben}{\begin{enumerate}}
\nc{\een}{\end{enumerate}}
\nc{\bfig}{\begin{figure}}
\nc{\efig}{\end{figure}}
\nc{\arreq}{&\!=\!&}
\nc{\arrmi}{&\!-\!&}
\nc{\arrpl}{&\!+\!&}
\nc{\arrap}{&\!\!\!\approx\!\!\!&}
\nc{\non}{\nonumber\\*}
\nc{\align}{\!\!\!\!\!\!\!\!&&}

\def\lsim{\; \raise0.3ex\hbox{$<$\kern-0.75em
      \raise-1.1ex\hbox{$\sim$}}\; }
\def\gsim{\; \raise0.3ex\hbox{$>$\kern-0.75em
      \raise-1.1ex\hbox{$\sim$}}\; }
\nc{\DOT}{\hspace{-0.08in}{\bf .}\hspace{0.1in}}
\nc{\Laada}{\hbox {$\sqcap$ \kern -1em $\sqcup$}}
\nc\loota{{\scriptstyle\sqcap\kern-0.55em\hbox{$\scriptstyle\sqcup$}}}
\nc\Loota{{\sqcap\kern-0.65em\hbox{$\sqcup$}}}
\nc\laada{\Loota}
\nc{\qed}{\hskip 3em \hbox{\BOX} \vskip 2ex}

\nc{\real}{{\rm I \! R}}
\nc{\Z}{{\sf Z \!\!\! Z}}
\nc{\complex}{{\rm C\!\!\! {\sf I}\,\,}}
\def\bigid{\leavevmode\hbox{\small1\kern-3.8pt\normalsize1}}
\def\id{\leavevmode\hbox{\small1\kern-3.3pt\normalsize1}}
\nc{\slask}{\!\!\!/}
\nc{\bis}{{\prime\prime}}
\nc{\pa}{\partial}
\nc{\na}{\nabla}
\nc{\ra}{\rangle}
\nc{\la}{\langle}
\nc{\goto}{\rightarrow}
\nc{\swap}{\leftrightarrow}

\nc{\EE}[1]{ \mbox{$\cdot10^{#1}$} }
\nc{\abs}[1]{\left|#1\right|}
\nc{\at}[2]{\left.#1\right|_{#2}}
\nc{\norm}[1]{\|#1\|}
\nc{\abscut}[2]{\Abs{#1}_{\scriptscriptstyle#2}}
\nc{\vek}[1]{{\rm\bf #1}}
\nc{\integral}[2]{\int\limits_{#1}^{#2}}
\nc{\inv}[1]{\frac{1}{#1}}
\nc{\dd}[2]{{{\partial #1}\over{\partial #2}}}
\nc{\ddd}[2]{{{{\partial}^2 #1}\over{\partial {#2}^2}}}
\nc{\dddd}[3]{{{{\partial}^2 #1}\over
        {\partial #2 \partial #3}}}
\nc{\dder}[2]{{{d #1}\over{d #2}}}
\nc{\ddder}[2]{{{d^2 #1}\over{d {#2}^2}}}
\nc{\dddder}[3]{{d^2 #1}\over
        {d #2 d #3}}
\nc{\dx}[1]{d\,^{#1}x}
\nc{\dy}[1]{d\,^{#1}y}
\nc{\dz}[1]{d\,^{#1}z}
\nc{\dl}[1]{\frac{d\,^{#1}l}{(2\pi)^{#1}}}
\nc{\dk}[1]{\frac{d\,^{#1}k}{(2\pi)^{#1}}}
\nc{\dq}[1]{\frac{d\,^{#1}q}{(2\pi)^{#1}}}

\nc{\cc}{\mbox{$c.c.$ }}
\nc{\hc}{\mbox{$h.c.$ }}
\nc{\cf}{cf.\ }
\nc{\erfc}{{\rm erfc}}
\nc{\Tr}{{\rm Tr\,}}
\nc{\tr}{{\rm tr\,}}
\nc{\pol}{{\rm pol}}
\nc{\sign}{{\rm sign}}
\nc{\bfT}{{\bf T }}

\def\keV{{\rm\ keV}}

\nc{\cA}{{\cal A}}
\nc{\cB}{{\cal B}}
\nc{\cD}{{\cal D}}
\nc{\cE}{{\cal E}}
\nc{\cG}{{\cal G}}
\nc{\cH}{{\cal H}}
\nc{\cL}{{\cal L}}
\nc{\cO}{{\cal O}}
\nc{\cT}{{\cal T}}
\nc{\cN}{{\cal N}}
\nc{\rvac}[1]{|{\cal O}#1\rangle}
\nc{\lvac}[1]{\langle{\cal O}#1|}
\nc{\rvacb}[1]{|{\cal O}_\beta #1\rangle}
\nc{\lvacb}[1]{\langle{\cal O}_\beta #1 |}
\nc{\bb}{\bar{\beta}}
\nc{\bt}{\tilde{\beta}}
\nc{\ctH}{\tilde{\cal H}}
\nc{\chH}{\hat{\cal H}}
\nc{\1}{\aa}
\nc{\2}{\"{a}}
\nc{\3}{\"{o}}
\nc{\4}{\AA}
\nc{\5}{\"{A}}
\nc{\6}{\"{O}}
\nc{\al}{\alpha}
\nc{\g}{\gamma}
\nc{\Del}{\Delta}
\nc{\e}{\epsilon}
\nc{\eps}{\epsilon}
\nc{\lam}{\lambda}
\nc{\om}{\omega}
\nc{\Om}{\Omega}
\nc{\ve}{\varepsilon}
\nc{\mn}{{\mu\nu}}
\nc{\k}{\kappa}
\nc{\vp}{\varphi}

\nc{\advp}[3]{{\it  Adv.\ in\ Phys.\ }{{\bf #1} {(#2)} {#3}}}
\nc{\annp}[3]{{\it  Ann.\ Phys.\ (N.Y.)\ }{{\bf #1} {(#2)} {#3}}}
\nc{\apl}[3]{{\it  Appl. Phys. Lett. }{{\bf #1} {(#2)} {#3}}}
\nc{\apj}[3]{{\it  Ap.\ J.\ }{{\bf #1} {(#2)} {#3}}}
\nc{\apjl}[3]{{\it  Ap.\ J.\ Lett.\ }{{\bf #1} {(#2)} {#3}}}
\nc{\app}[3]{{\it Astropart.\ Phys.\ }{{\bf #1} {(#2)} {#3}}}
\nc{\cmp}[3]{{\it  Comm.\ Math.\ Phys.\ }{{ \bf #1} {(#2)} {#3}}}
\nc{\cqg}[3]{{\it  Class.\ Quant.\ Grav.\ }{{\bf #1} {(#2)} {#3}}}
\nc{\epl}[3]{{\it  Europhys.\ Lett.\ }{{\bf #1} {(#2)} {#3}}}
\nc{\ijmp}[3]{{\it Int.\ J.\ Mod.\ Phys.\ }{{\bf #1} {(#2)} {#3}}}
\nc{\ijtp}[3]{{\it Int.\ J.\ Theor.\ Phys.\ }{{\bf #1} {(#2)} {#3}}}
\nc{\jmp}[3]{{\it  J.\ Math.\ Phys.\ }{{ \bf #1} {(#2)} {#3}}}
\nc{\jpa}[3]{{\it  J.\ Phys.\ A\ }{{\bf #1} {(#2)} {#3}}}
\nc{\jpc}[3]{{\it  J.\ Phys.\ C\ }{{\bf #1} {(#2)} {#3}}}
\nc{\jap}[3]{{\it J.\ Appl.\ Phys.\ }{{\bf #1} {(#2)} {#3}}}
\nc{\jpsj}[3]{{\it J.\ Phys.\ Soc.\ Japan\ }{{\bf #1} {(#2)} {#3}}}
\nc{\lmp}[3]{{\it Lett.\ Math.\ Phys.\ }{{\bf #1} {(#2)} {#3}}}
\nc{\mpl}[3]{{\it  Mod.\ Phys.\ Lett.\ }{{\bf #1} {(#2)} {#3}}}
\nc{\ncim}[3]{{\it  Nuov.\ Cim.\ }{{\bf #1} {(#2)} {#3}}}
\nc{\np}[3]{{\it  Nucl.\ Phys.\ }{{\bf #1} {(#2)} {#3}}}
\nc{\pr}[3]{{\it Phys.\ Rev.\ }{{\bf #1} {(#2)} {#3}}}
\nc{\pra}[3]{{\it  Phys.\ Rev.\ A\ }{{\bf #1} {(#2)} {#3}}}
\nc{\prb}[3]{{\it  Phys.\ Rev.\ B\ }{{{\bf #1} {(#2)} {#3}}}}
\nc{\prc}[3]{{\it  Phys.\ Rev.\ C\ }{{\bf #1} {(#2)} {#3}}}
\nc{\prd}[3]{{\it  Phys.\ Rev.\ D\ }{{\bf #1} {(#2)} {#3}}}
\nc{\prl}[3]{{\it Phys.\ Rev.\ Lett.\ }{{\bf #1} {(#2)} {#3}}}
\nc{\pl}[3]{{\it  Phys.\ Lett.\ }{{\bf #1} {(#2)} {#3}}}
\nc{\prep}[3]{{\it Phys\. Rep.\ }{{\bf #1} {(#2)} {#3}}}
\nc{\prsl}[3]{{\it Proc.\ R.\ Soc.\ London\ }{{\bf #1} {(#2)} {#3}}}
\nc{\ptp}[3]{{\it  Prog.\ Theor.\ Phys.\ }{{\bf #1} {(#2)} {#3}}}
\nc{\ptps}[3]{{\it  Prog\ Theor.\ Phys.\ suppl.\ }{{\bf #1} {(#2)} {#3}}}
\nc{\physa}[3]{{\it  Physica\ A\ }{{\bf #1} {(#2)} {#3}}}
\nc{\physb}[3]{{\it  Physica\ B\ }{{\bf #1} {(#2)} {#3}}}
\nc{\phys}[3]{{\it Physica\ }{{\bf #1} {(#2)} {#3}}}
\nc{\rmp}[3]{{\it  Rev.\ Mod.\ Phys.\ }{{\bf #1} {(#2)} {#3}}}
\nc{\rpp}[3]{{\it Rep.\ Prog.\ Phys.\ }{{\bf #1} {(#2)} {#3}}}
\nc{\sjnp}[3]{{\it Sov.\ J.\ Nucl.\ Phys.\ }{{\bf #1} {(#2)} {#3}}}
\nc{\spjetp}[3]{{\it Sov.\ Phys.\ JETP\ }{{\bf #1} {(#2)} {#3}}}
\nc{\yf}[3]{{\it Yad.\ Fiz.\ }{{\bf #1} {(#2)} {#3}}}
\nc{\zetp}[3]{{\it Zh.\ Eksp.\ Teor.\ Fiz.\  }{{\bf #1}  {(#2)} {#3}}}
\nc{\zp}[3]{{\it Z.\ Phys.\ }{{\bf #1} {(#2)} {#3}}}
\nc{\ibid}[3]{{\sl ibid.\ }{{\bf #1} {#2} {#3}}}
\nc{\rf}[1]{(\ref{#1})}
\nc{\nn}{\nonumber \\*}
\nc{\bfB}{\bf{B}}
\nc{\bfv}{\bf{v}}
\nc{\bfx}{\bf{x}}
\nc{\bfy}{\bf{y}}
\nc{\vx}{\vec{x}}
\nc{\vy}{\vec{y}}
\nc{\oB}{\overline{B}}
\nc{\oI}{\overline{I}}
\nc{\oR}{\overline{R}}
\nc{\rar}{\rightarrow}
\nc{\ti}{\times}
\nc{\slsh}{\hskip-5pt/}
\nc{\sm}{Standard~Model~}
\nc{\MP}{M_{\rm Pl}}
\nc{\tp}{t_{\rm Pl}}
\nc{\ave}{\bar{E}}

\nc{\eff}{{\rm eff}}
\nc{\kk}{\vek{k}}
\nc{\pp}{{\rm p}}
\nc{\ga}{g_{a\gamma}}
\nc{\vv}{\\}
\nc{\eee}{{\bf E}}
\nc{\bbb}{{\bf B}}
\nc{\qcd}{T_{\rm QCD}}
\nc{\G}{\rm \ G}
\def\vec#1{{\bf #1}}
\begin{document}
 \title
{\bf Cosmological abundances of right-handed Dirac neutrinos}
\author{
{\sc Kari Enqvist}}
\address{\sl Department of Physics\\ 
University of Helsinki, FIN-00014 Helsinki, Finland}
\maketitle
\abstracts{The equilibration of the right-helicity
states $\nu_+$ of light Dirac-neutrinos is discussed. I point out
that the  $\nu_+$ production rate is enhanced by weak gauge boson
pole effects so that the right-helicity component of $\nu_\tau$
is brought into equilibrium at $T\simeq 10$ GeV independently
of the initial abundance, provided $m_{\nu_{\tau}}\gsim 10$ keV.
Neutrino spin flip in primordial magnetic fields and the
resulting bound on $\mu_\nu$ is also
discussed.
}

%
\section{Relic abundances of right-helicity neutrinos}
Primordial nucleosynthesis is a remarkable probe of neutrino properties
\cite{subir}.  To some extent
primordial nucleosynthesis could be sensitive even
to the Dirac vs. Majorana  nature of neutrinos, because in the Dirac case
the small relic abundance of the inert right-handed component of
Dirac neutrino would also contribute at nucleosynthesis. Of course,
presently one cannot  hope to  differentiate between the
Dirac and Majorana nature of neutrinos on cosmological grounds, but
in principle this is an interesting problem. 
The right-helicity states of Dirac neutrinos can be
produced (and destroyed) in spin-flip transitions induced by the Dirac mass
\cite{mass,kimmo} or the neutrino magnetic moment. Spin-flip transitions
may also be induced by primordial magnetic fields, if such exist.

The actual cosmological density of the right-helicity neutrinos depends
not only on the production rate near the QCD phase transition, but also
on whether the right-helicity neutrinos had a chance to equilibrate
at some point during the course of the evolution of the universe.
This depends on the thermal scattering rates of neutrinos.
An important
source of $\nu_+$'s are also the non-equilibrium neutrino scatterings and
decays of pions \cite{kimmo}. Such processes give rise to the  bound
\cite{FieldsKimmoOliivi}
$m_{\nu_{\mu}} \lsim 130 \keV$ and $m_{\nu_{\tau}} \lsim 150 \keV$, 
using $T_{\rm QCD}=100$~MeV and assuming
that  nucleosynthesis allows  less than $0.3$ extra neutrino
families. Let us note that there seems to be no window of opportunity 
\cite{DolgovPastorValle} for a sufficiently 
stable ($\tau_\nu\gsim 10^2$~sec) tau neutrino
in the MeV region because of the production of non-equilibrium electron
neutrinos in $\nu_\tau\bar{\nu}_\tau$ annihilations,
 which would disrupt the succesful nucleosynthesis
predictions.

In the Standard Model
there are \cite{ekmu}  68 purely fermionic
$2\to 2$ processes  in which a right-helicity
muon or tau neutrino can be produced. In addition,
a right-helicity tau neutrino can also be produced in  11 lepton and quark
three-body
decays, and the  muon neutrino in another set of 11 three-body
decays. In principle, one has also to consider processes involving
$W^{\pm}$, $Z$ and $H$. There are
 16  such processes.
Finally, there are 3 two-body decays  of $W^{\pm}$, $Z$
and $H$ bosons which are capable producing right-helicity muon and tau
neutrinos. 
 
The thermally averaged production rate in $2\to 2$ scattering 
$a+b\to \nu_{+}+d$ per one right-helicity neutrino $\nu_+$ is
\bea{rate}
\Gamma_+ & = &  \frac{1}{n^{\rm FD}_{+}}
           \int
           d\Pi_a d\Pi_b d\Pi_+ d\Pi_d (2\pi )^4
           \delta^{(4)} (p_a + p_b - p_+ - p_d)
               S |{\cal M}_{ab\to +d}|^2 \nonumber \\
           & & \quad\quad\quad\quad \times f_a^{\rm FD} f_b^{\rm FD}
               (1-f_+^{\rm FD}) (1-f_d^{\rm FD})~,
\eea
where $n^{\rm FD}_{+}$ is the the equilibrium
number density of the right-handed neutrinos, $d\Pi_i \equiv d^3p_i/((2\pi)^3
2E_i)$,  $S$
is the symmetry factor taking into account identical particles in the initial
and/or final
states, and $f_i^{\rm FD}$ are Fermi-Dirac distribution functions.
The fermionic processes exhibit an enhancement
of the $\nu_+$ production rate which is due to gauge boson pole effects
\cite{ekmu}. This is demonstrated in Fig. 1, where the enhancement is apparent
in the s-channel process $u\overline d\to\nu_{+}\tau^+$, where $\nu_+$ is a
right helicity tau neutrino, as compared with the crossed t-channel process.
It turns out that in the temperature of interest,
the processes involving gauge or Higgs bosons can  be neglected in comparision
with the purely fermionic processes.
This is demonstrated in Fig. 1, where  the thermally averaged rate
for the bosonic process $\tau^-\gamma \rightarrow \nu_{+} W^-$ is shown.
\begin{figure}
\leavevmode
\centering
\vspace*{75mm}
\includegraphics{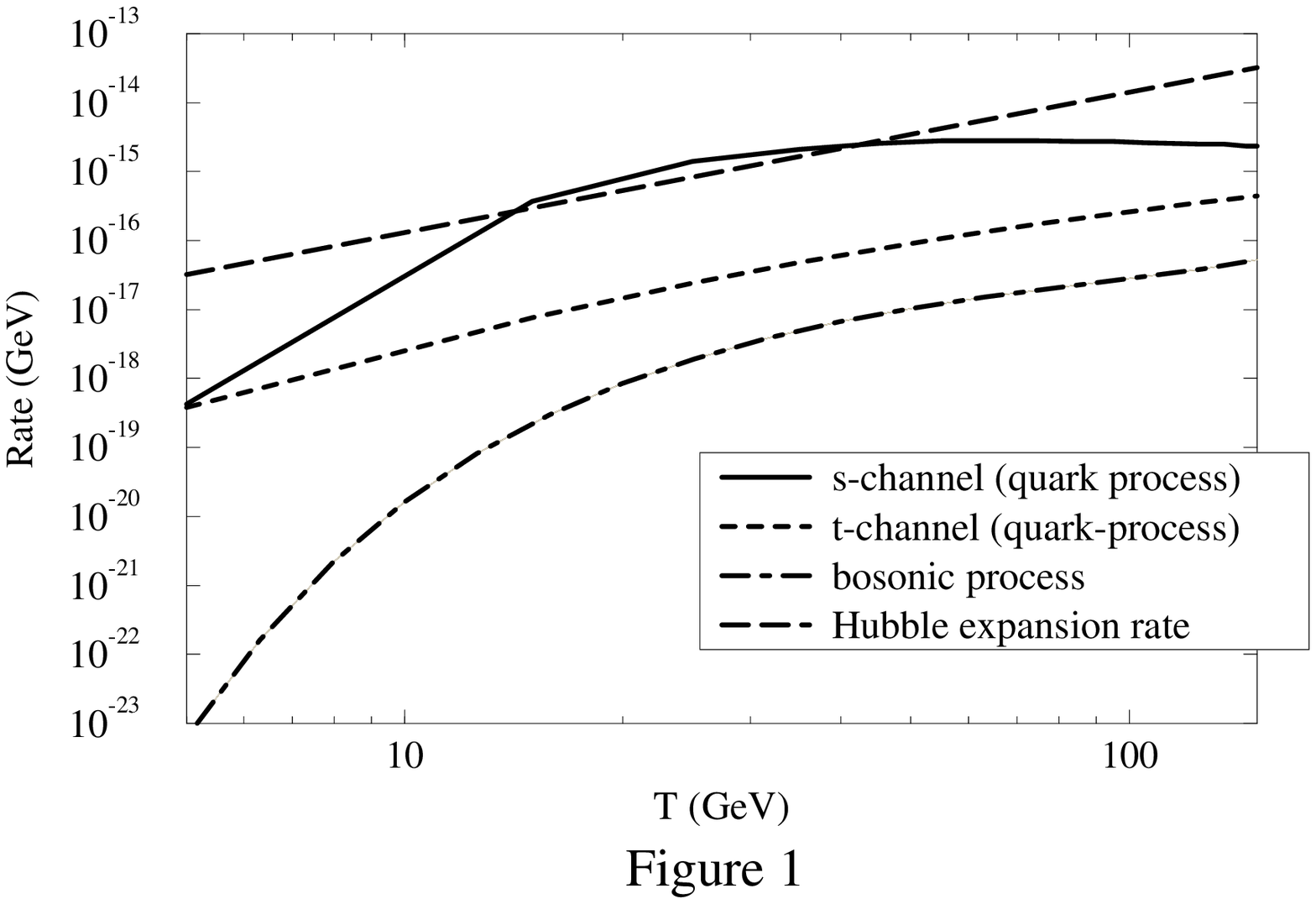}
\caption{Thermally averaged production rate for the 
s-channel reaction
$u\overline d\to\nu_{+}\tau^+$ with $m_{\nu_{\tau}}=20$ keV; 
shown is also the corresponding t-channel
rate as well as the rate for 
the bosonic process $\tau^-\gamma \rightarrow \nu_{+} W^-$.}
\label{kuvaaxel2}
\end{figure}

Let us note that at high $T$ the rate \eq{rate} 
is infrared sensitive to the thermal
corrections in the propagators. In most cases
it is an excellent approximation just to modify the propagators by
introducing a Debye mass $M^2(T)=\Pi_L(\omega, \kv=0)$, which may
be approximated by
$M_i^2(T)\simeq M_i^2 + 0.1\; T^2$ ($i=W,Z$). 

The relic density of the right-helicity tau neutrinos can be 
found by solving the Boltzmann equation
\begin{equation}
  \left( \frac{\partial}{\partial t}
         -H|{\pv}_+|\frac{\partial}{\partial |{\pv}_+|} \right) f_+
  = \left( \frac{\partial f_+}{\partial t} \right)_{\rm coll}~.
  \label{BoltzEq:t}
\end{equation}
Considering only purely fermionic $2\to 2$ and $1\to 3$ processes, 
we can write
the collision term  in the form
\begin{equation}
  \left( \frac{\partial {f}_+}{\partial R} \right)_{\rm coll} =
    (C_{2\to 2}+C_{1\to 3})(1-{f}_+)
  - (C_{2\to 2}^\prime + C_{1\to 3}^\prime) {f}_+~,
  \label{RHS:Cs}
\end{equation}
where  the coefficients $C_I$ represent production and
$C'_I$ destruction of $\nu_+$, with $I = 2\to 2, 1\to 3$.
The explicit expressions for the quantities $C_{2\to 2}$ and $C_{1\to 3}$
are
\begin{eqnarray}
  C_{2\to 2}(|{\pv}_{+}|,R) & = &\sum_{\rm scatt} \frac{1}{2E_+}
    \int d\Pi_a d\Pi_b d\Pi_d (2\pi)^4 \delta^{(4)}(p_a+p_b-p_{+}-p_d)
    \nonumber\\
    & & \quad\quad\quad\quad\quad\quad\quad\times\, S|{\cal
M}_{ab\leftrightarrow {+}d}|^2
    f_a^{\rm FD} f_b^{\rm FD} (1-f_d^{\rm FD})~, \nonumber\\
  C_{1\to 3}(|{\pv}_{+}|,R) & = &\sum_{\rm dec} \frac{1}{2E_+}
    \int d\Pi_f d\Pi_g d\Pi_h (2\pi)^4 \delta^{(4)}(p_f-p_g-p_{+}-p_h)
    \nonumber\\
    & & \quad\quad\quad\quad\quad\quad\quad \times\, S|{\cal
M}_{f\leftrightarrow g{+}h}|^2
    f_f^{\rm FD} (1-f_g^{\rm FD}) (1-f_h^{\rm FD})~.
    \label{Expr:Cs}
\end{eqnarray}

The Boltzmann equation \eq{BoltzEq:t}
can be solved numerically (by using e.g. 30 momentum 
bins \cite{ekmu}).  One can consider
two extreme initial conditions for $\nu_+$ at $T\simeq 100$ GeV: {\it (i)}
full equilibrium with $f_+=f_{\rm eq}$
and {\it (ii)} complete decoupling with $f_+=0$. The evolution of
the  $\nu_+$  energy density $\rho_+$ for different masses 
and for the two different initial conditions is shown in
Fig. 2. One observes that right-helicity tau neutrinos equilibrate 
independently of the initial condition provided $m_{\nu_{\tau}}\gsim 10$
keV. The relic density, in units of extra neutrino species
$\Delta N_\nu$, is also shown in Fig. 2. 
For $m_{\nu_{\tau}}\lsim 10$ keV the relic density is very close to the
expected value $\rho_+/\rho_L=[g_*(T)/g_*(T_i)]^{4/3}\rho_+(T_i)/\rho_L(T_i)$,
where $\rho_L$ is the density of left-handed neutrinos
and $T_i$ is the initial temperature.
For $m_{\nu_{\tau}}\gsim 10$ keV one would obtain a small extra effect at
nucleosynthesis, independently of the initial abundance.
The results are for $\nu_\tau$, but for 
$\nu_\mu$ the graphs would look very similar.
\begin{figure}
\leavevmode
\centering
\vspace*{75mm}
\includegraphics{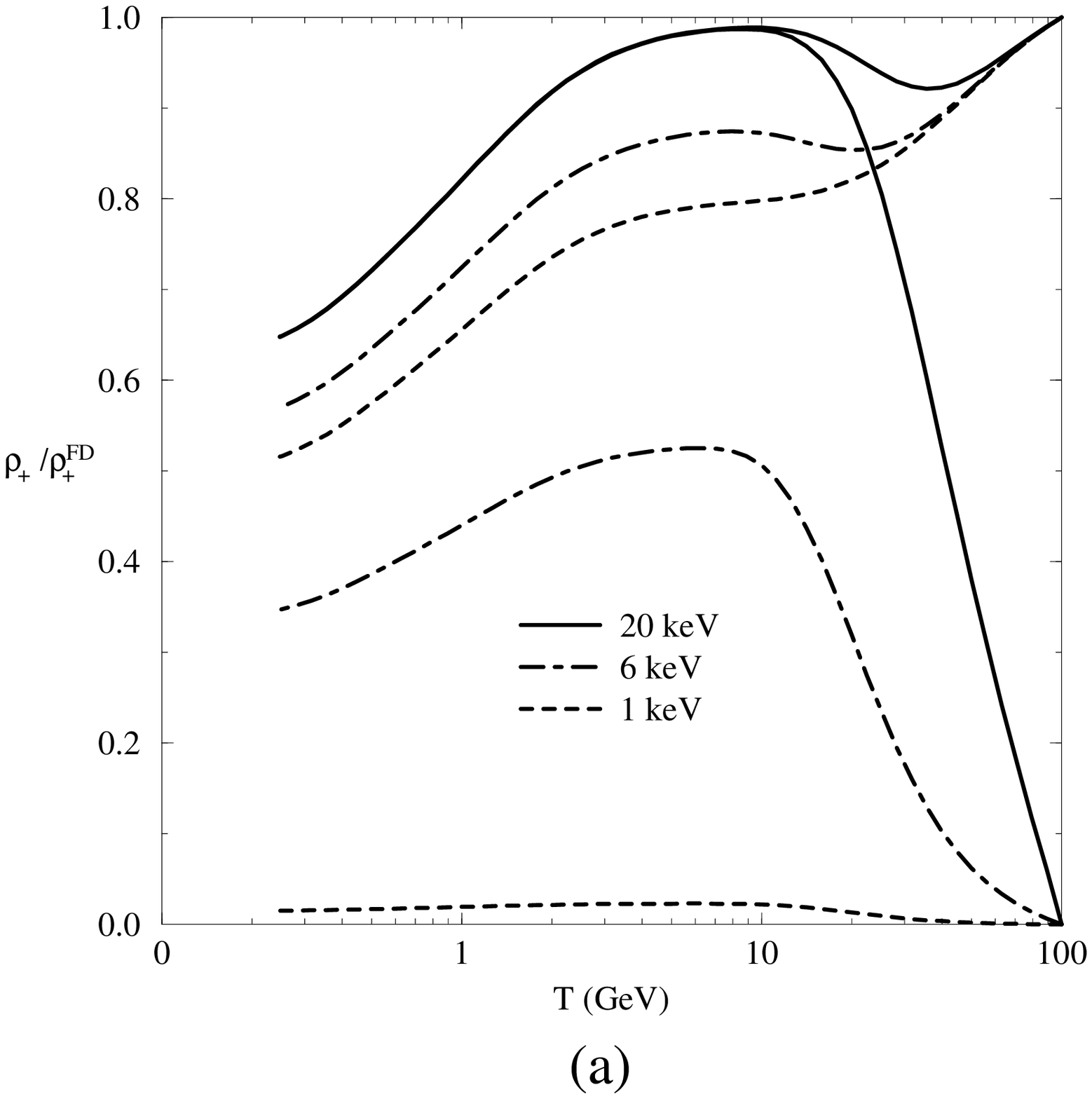}
\includegraphics{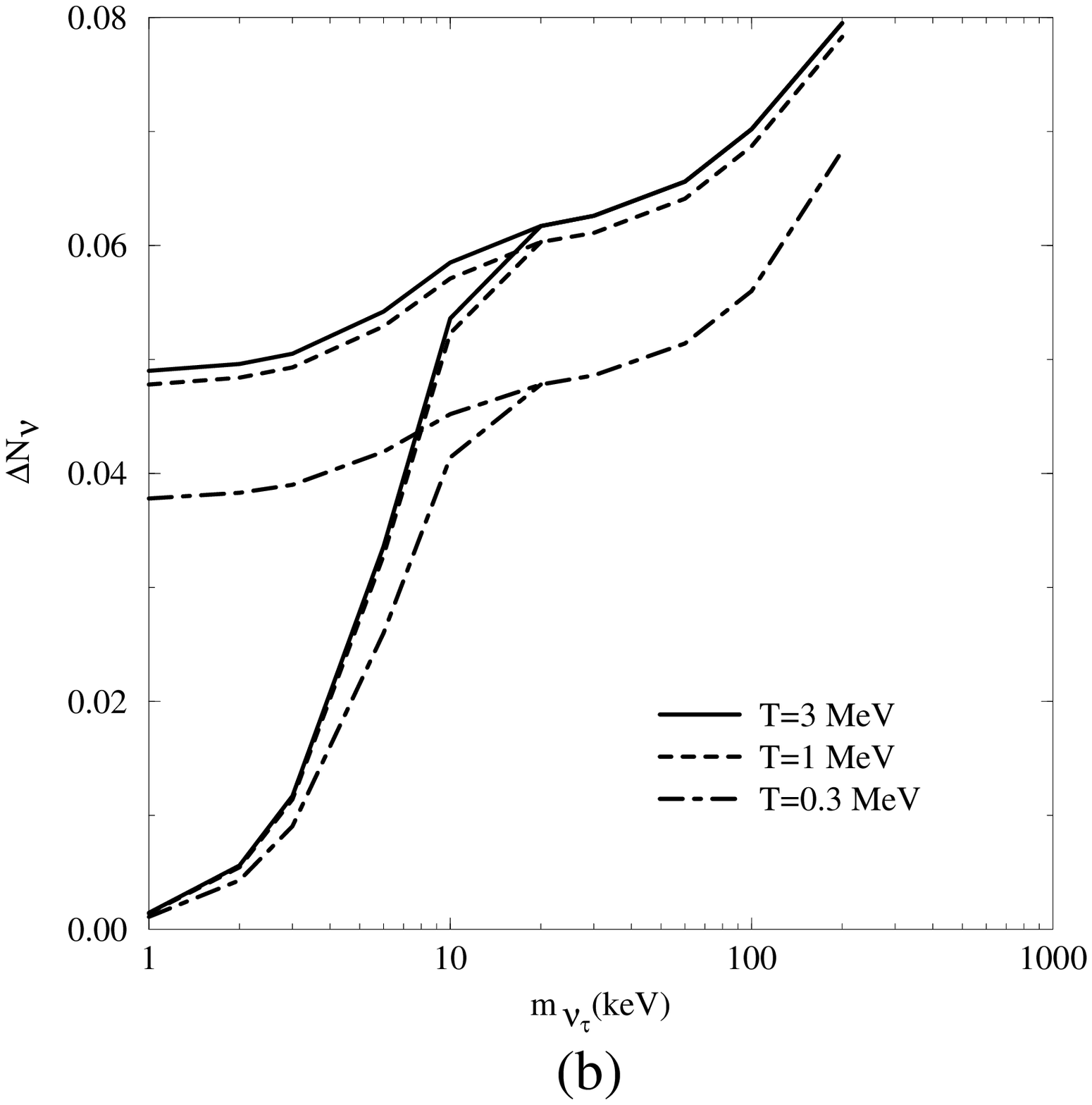}
\caption{(a) The evolution of the  right-helicity tau neutrino
relic density $\rho_+$, in units
of the equilibrium density $\rho^{\rm FD}$; (b) the relic density
of the  right-helicity tau neutrino in units of extra neutrino species; 
the upper set of curves is for right-helicity neutrinos
in equilibrium at $T\simeq 100$ GeV, the lower set corresponds to
zero initial density.}
\label{kuvaaxel3}
\end{figure}
\section{Effects of magnetic fields}
The abundance of the right-helicity states is also sensitive to 
primordial magnetic fields. Large magnetic fields could be created by 
cosmological phase transitions at microscopic length scales 
\cite{largeB,poul}. 
There are indications that the small scale field 
cascades into larger length scales because of hydromagnetic turbulence
\cite{cascade}.
Electrical conductivity
in the early universe is high \cite{conduct} so that the field is frozen
in the plasma. Typically, such a primordial magnetic field is random
so that $\langle{\bf B}\rangle=0$.
When a Dirac neutrino propagates through such magnetic field, its spin
will precess. At the same time, the neutrino is subject to scattering
and thermal corrections. Neutrino dispersion relations in magnetized
plasma have recently been carefully studied \cite{elmfors}.
The actual spin evolution is best described by a relativistic
kinetic equation \cite{RKE}, and the result depends very much on whether
the coherence length of ${\bf B}$ is larger or smaller than the neutrino
scattering length. Another unknown is how to average over the randomly
varying magnetic field. In the case of neutrino propagation, the appropriate
statistical procedure might be the line average \cite{poul}.
Purely phenomenologically, one can write
\begin{equation}
B_{\rm rms}=B_0\left({T\over T_0}\right)^2\left({d\over L_0}\right)^p~,
\label{rootmean}
\end{equation}
where $L_0$ is the coherence length of the magnetic field, $d$ is the
distance scale and $p$ is unknown; for a line average, $p=1/2$.

In the case of small-scale random magnetic field, 
neutrino forward scattering tends to depolarize the spin \cite{ers}.
If the field is completely uncorrelated at distances $d\gg L_0$ so that
\begin{equation}
\langle B(t)B(t_1)\rangle=B^2_{\rm rms}L_0\delta(t-t_1)~,
\label{correlator}
\end{equation}
one finds that the spin flip probability reads \cite{ers}
\begin{equation}
P_{\nu_L\to\nu_R}=\frac12 \left(1-{\rm exp}(-\Gamma t)\right)~,
\label{correlator2}
\end{equation}
where the damping parameter $\Gamma$ is given by
\begin{equation}
\Gamma=\frac83\mu_\nu^2B^2_{\rm rms}L_0^2~.
\label{damp}
\end{equation}
Setting $t=H^{-1}$ and requiring that $\Gamma<H$ at $T=T_{\rm QCD}$
so that the right-helicity states are not in equilibrium below QCD
phase transition, and in particular during nucleosynthesis, 
results in the bound
\begin{equation}
\mu_\nu B_{\rm rms}(T_{\rm QCD},H^{-1})\lsim 6.7\times10^{-3}\mu_BG
\left({L_W\over L_0}\right)^{1/2}~.
\label{raja}
\end{equation}
In order to translate this to a bound on $\mu_\nu$ one needs to assume
something about the magnitude of $B_{\rm rms}$ at the horizon
scale at $T=T_{\rm QCD}$. No reliable estimate exists at present time.
However, the bound \eq{raja} could in principle be as restrictive
as \cite{ers} $\mu_\nu\lsim 10^{-20}\mu_B$.

\end{document}